\documentclass[aps,prb,showpacs,superscriptaddress,twocolumn]{revtex4}
\usepackage{graphicx,psfrag}

\begin{document}
\title{Width of reaction zones in $A+B\rightarrow C$ type 
reaction-diffusion processes: \\
Effects of an electric current}

\author{K. Martens}
\affiliation{Department of Theoretical Physics, 
University of Gen\`eve, CH-1211 Gen\`eve 4, Switzerland}
\author{M. Droz}
\affiliation{Department of Theoretical Physics,
University of Gen\`eve, CH-1211 Gen\`eve 4, Switzerland}
\author{Z. R\'acz}
\affiliation{Institute for Theoretical Physics - 
HAS, E\"otv\"os University, P\'azm\'any s\'et\'any 1/a, 1117 Budapest, Hungary}

\date{\today}

\begin{abstract}

We investigate the effects of an electric current 
on the width
of a stationary reaction zone in an irreversible $A^- +B^+ \to C$
reaction-diffusion process. The ion dynamics 
of the electrolytes
$A\equiv (A^+,A^-)$ and $B\equiv (B^+,B^-)$ is described by
reaction-diffusion equations obeying local electroneutrality, and
the stationary state is obtained by employing reservoirs
of fixed electrolyte concentrations at the
opposite ends of a finite domain. We find that
the width of the reaction zone decreases when 
the current drives the reacting ions towards the reaction zone 
while it increases
in the opposite case. The linear response of the width to 
the current is estimated by developing
a phenomenological theory based on conservation laws and on 
electroneutrality. 
The theory is found to reproduce 
numerical solutions to a good accuracy.

\end{abstract}
\pacs{05.60.Gg, 64.60.Ht, 75.10.Jm, 72.25.-b}
\maketitle

\section{Introduction}
\label{introduction}

The reaction-diffusion process $A+B\rightarrow C$ 
combined with the ensuing evolution of the reaction product $C$ 
(e.g. precipitation) 
underlie the explanation of a large number of physical, 
chemical, and biological 
phenomena~\cite{Henisch,kotomin,Dani,mathbiol}.
An interesting aspect of the reaction-diffusion part of 
these processes is the presence of reaction zones. They are
formed either because the reagents $A$ and $B$ are initially 
separated \cite{zol} or because spatial inhomogeneities 
exist in the initial 
distribution of the reagents \cite{Redner1}. 
These zones are important since by 
determining where and when the reaction product emerges, 
they set the stage for the temporal and spatial evolution of $C$s. 
Accordingly, the motion of these fronts, the
spatial distribution of the rate of the production of $C$, and the
width of the reaction zones have been much investigated. They 
are known theoretically for the case of neutral 
reagents~\cite{zol,Redner1,chopard,cornell,inhom, redner} and
the theories have been verified in experiments \cite{Kopelman-1,Bazant,Kopelman-3,Tabeling,Kopelman-4}.

In realistic situations, however, the reagents $A$ and $B$ are  
often electrolytes which dissociate,
\begin{equation}
A \rightarrow A^++ A^-\,,\qquad B \rightarrow B^++B^-\,,
\label{dissociation}
\end{equation} 
and the reaction takes place between the `active' ions which,
for definiteness, will be taken below to be $A^-$ and $B^+$: 
\begin{equation}
A^-+B^+\rightarrow C \,.
\label{reaction}
\end{equation}
Although the counterions $A^+$ and $B^-$ are not reacting, they 
influence  the dynamics significantly through the 
electroneutrality constraint so that the task of characterizing 
the front becomes much more involved~\cite{rubinstein90}. 
 
Some of the front properties such as the spatial location and 
the reaction product distribution (but not the width of the 
reaction zone) have nevertheless been obtained for ionic reactions in 
one-dimensional geometry~\cite{unger}. Furthermore, these studies 
have been extended to ions being driven by an electric current
or by an external potential difference~\cite{bena}. The driven 
systems are of special interest when the $C$s undergo 
phase separation, since 
then they may be used to design  
bulk precipitation patterns. 
Indeed, it has been shown recently~\cite{ourprl,encoding} 
that a flexible control  
of precipitation patterns can be achieved through controlling the
reaction zones by
appropriately designed time-dependent currents.

From the technological point of view, controlling precipitation patterns
becomes relevant if the patterns can be downsized to the 
submicron range. Since the width of the reaction zones 
is one of the limiting factors in 
downsizing, it is clear that one should understand how to 
control it. The studies of the width for neutral reagents~\cite{zol,chopard,cornell,redner}
suggest that the parameter strongly affecting the width is the 
reaction rate constant. It is, however, not a parameter we can adjust,
thus other means of control should be found. Since 
electric currents turned out to be useful in manipulating
patterns~\cite{ourprl,encoding}, it is natural to ask if the width 
could also be controlled by them. This is the question we address
in this work.

In order to simplify the task, 
we restrict our study to a one-dimensional reaction-diffusion
process on a finite interval.
The concentrations of the electrolytes are fixed at the
boundaries and, furthermore, a 
current generator is attached so that a constant 
current flows through the system 
(see Fig.\ref{fig1} and Fig.\ref{fig2}). For this setup, we derive 
the reaction-diffusion equations in the 
long-time limit when the stationary state is reached.
Solving these
equations numerically (and, in some limits, analytically as well),
we obtain the width of the stationary reaction front, $w(J)$, 
as a function of the electric current, $J$. 
 
The $J=0$ case has been investigated earlier \cite{redner} and
we recover the zero current width, $w(0)$, obtained in that work.
For $J\not= 0$, our general finding is that a 
{\it forward current} (a current 
that drives the reacting ions towards the reaction zone) reduces 
the width of the stationary reaction zone, while a current 
of opposite polarity (backward current) increases the width. 
For small currents the change compared to the zero-current case is 
proportional to the current [$w(J)-w(0)\propto J$] 
with a proportionality constant
that can be estimated from rather general reasoning and the 
results are found to compare favorably with the numerical solutions.
Although our results concern the width of stationary states, 
we present some arguments that 
the quasistationary nature of the diffusive fronts allows
the derivation of the dynamics of the width in some time window 
in case of moving fronts as well. 

The paper is organized as follows. In Sec.~II, we introduce 
the problem and discuss the equations describing the 
dynamics of the ions. Sec.~III contains the derivation of the 
model equations for the particular
case of the stationary state. 
Analytical and approximate solutions are obtained both for 
symmetric setup (Sec.~IV) and 
asymmetric setup (Sec.~V). Possible applications of the stationary
state results to moving reaction front are presented in Sec.~VI, 
and conclusions are drawn in Sec.~VII.

\section{The Problem and the model}

\begin{figure}[htbp]
\includegraphics[width=7cm,angle=0,clip]{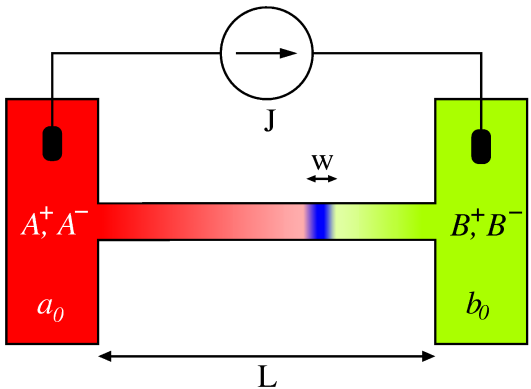}
\caption{(Color online) Experimental setup for generating 
a stationary reaction front in the presence of an electric 
current. The details are described in the text.}
\label{fig1}
\end{figure}

Fig.\ref{fig1} displays the setup for producing a stationary 
reaction zone in the presence of an electric current.
Two dissociating electrolytes $A$ and $B$ 
[see Eq.(\ref{dissociation})] are dissolved 
in a column of gel where their transport is restricted to be diffusive. 
The opposite ends of the columns are connected
to reservoirs of $A$ and $B$, respectively, thus we fix 
the electrolyte concentrations $a_0$ and $b_0$ at the boundaries. 
Furthermore, we maintain a constant electric current
through the system by attaching 
a current generator to the ends of the column separated 
by a distance $L$ (in another possible
setup one keeps a fixed potential difference between the ends).

\begin{figure}[htbp]
\includegraphics[width=7cm,angle=0,clip]{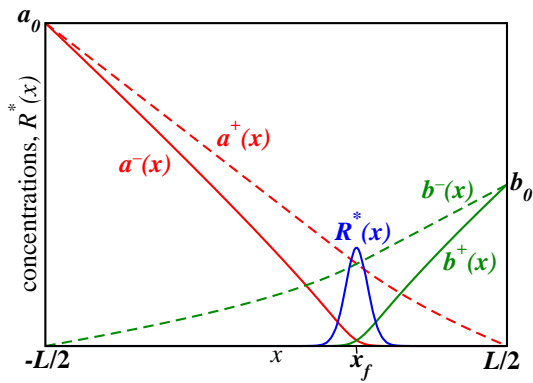}
\caption{(Color online) Characteristic stationary 
profiles of the ion 
concentrations in the presence of an electric current. 
The rate of production $R(x)$ of $C$s is measured 
in units of $ka_0$ where $k$ is the rate constant of the 
reaction, and a magnification factor $10^3$ is used 
for better visibility, i.e. $R^*=10^3R(x)/ka_0$. 
The position of the front is denoted by $x_f$.}
\label{fig2}
\end{figure}

As the reaction-diffusion dynamics proceeds,
a reaction zone (a spatial region where the reaction 
$A^-+B^+\to C$ takes place i.e. where the rate of the production 
$R(x,t)$ of $C$s is nonzero, see Fig.\ref{fig2}.) is formed. 
In the long-time limit the system relaxes to a stationary state 
and the front becomes stationary as well, $R(x,t)\to R(x)$. 
We shall be interested 
in the stationary state properties of the front, and 
particularly in its width $w$ which we can define e.g. 
through the second moment of the production rate  
\begin{equation}
w^2=\frac{\int \,dx\,(x-x_f)^2\,R(x)}{\int\,dx\,R(x)}
\label{width-def}
\end{equation}
where the integrals are over the interval $[-L/2,L/2]$ and 
$x_f$ is the center of the reaction zone
\begin{equation}
x_f=\frac{\int \,dx\,x\,R(x)}{\int\,dx\,R(x)}\,.
\label{xf-def}
\end{equation}

In order to calculate $w$, we shall use reaction-diffusion
equations which amounts to a mean-field description of the problem.
In addition, the following simplifying assumptions will be made:

(i) The system is treated as {\em one-dimensional},
i.e. all the relevant quantities are assumed to depend 
on a single spatial coordinate $x$, with
$-L/2\leq x \leq L/2$ being the spatial extent of the system.

(ii) We consider instantaneous 100\% {\em dissociation} of the electrolytes
$A$ and $B$ according to Eq.~(\ref{dissociation}) (this is the 
assumption of ``ideally strong" acid and basis). Accordingly, 
only $A^{\pm}$ and $B^{\pm}$ ions
(and the reaction product $C$) are present in the system.

(iii) The dynamics of the {\em inert} reaction product 
$C$ is assumed to have no feedback on the dynamics of the reagents. 

(iv) The equations are assumed to satisfy 
{\em electroneutrality}~\cite{rubinstein90} meaning that the local 
charge density is zero on relevant space scales ( in particular 
on the scale of the width of the reaction zone), i.e. 
\begin{equation}
\sum_i z_i \,n_i(x,t) =0\,,
\label{electroneutrality}
\end{equation}
where $z_i$ is the charge of the $i$-th ion -- in terms of the elementary charge $q$ -- and $n_i(x,t)$
denotes its density.  

(v) Monovalent ions, i.e. 
$|z_i|=1$ for all $i$ will be considered.

(vi) We assume {\em equal diffusion coefficients} for the ions, i.e. 
$D_i=D$ for all $i$.

(vii) The boundary conditions follow from the assumption that there 
are infinite reservoirs providing
fixed concentrations of the ions at the border . 
Denoting the concentration
profiles of the $A^{\pm}$ and $B^{\pm}$ ions 
by $a^{\pm}(x,t)$ and $b^{\pm}(x,t)$, respectively, the
boundary conditions then can be written as
\begin{eqnarray}
&&a^+(-L/2,t)=a^-(-L/2,t)={a_0}\,\label{bc1}\\
&&b^+(-L/2,t)=b^-(-L/2,t)=0\,\label{bc2}\\
&&a^+(L/2,t)=a^-(L/2,t)=0,\,\label{bc3}\\
&&b^+(L/2,t)=b^-(L/2,t)=b_0\,\label{bc4}.
\label{BC}
\end{eqnarray}
Since we are interested in the stationary state, the detailed initial 
conditions are irrelevant provided the reagents are separated initially.
Some dynamical properties, but not the width, 
have already been investigated for a similar setup~\cite{bena} for the case $a_0 
\ll b_0$ where the initial conditions lead to a moving front.

The general 
evolution equations for the concentration profiles
$a^{\pm}(x,t)$ and $b^{\pm}(x,t)$  were derived in~\cite{bena}. 
For the simplified case we are considering here, they 
acquire the following form:

\begin{eqnarray}
\frac{\partial a^-(x,t)}{\partial t}&=&D\frac{\partial^2 a^-}{\partial x^2}
+\lambda \frac{\partial (a^-{E})}{\partial x}-ka^-b^+
\label{electroneut1-0}\\
\frac{\partial b^+(x,t)}{\partial t}&=&D\frac{\partial^2 b^+}{\partial x^2} -\lambda \frac{\partial (b^+{E})}{\partial x}-ka^-b^+
\label{electroneut2-0}\\
\frac{\partial a^+(x,t)}{\partial t}&=&D\frac{\partial^2 a^+}{\partial x^2}-\lambda\frac{\partial (a^+ {E})}{\partial x}
\label{electroneut3-0}\\
\frac{\partial b^-(x,t)}{\partial t}&=&D\frac{\partial^2 b^-}{\partial x^2}+\lambda\frac{\partial (b^-{E})}{\partial x}\,.
\label{electroneut4-0}
\end{eqnarray}
Here $D$ are the
diffusion coefficients of the ions, $k$ is the rate constant of the 
reaction $A^{-}+B^{+}\to C$, $\lambda =DF/RT$ with $F=qN_A$ being 
the Faraday's constant 
(i.e., the electric charge transported by a mole
of monovalent positive ions), $R$ is the universal gas constant, 
and $T$ is the temperature.

The scaled local electric field $\lambda E(x,t)$ 
in the above equations is obtained from the 
{\em local electroneutrality} assumption, and 
it is given by~\cite{rubinstein90, bena}:
\begin{equation}
\lambda E(x,t)=
\frac{J(t)}
{q\,\left(\,a^++a^-+b^++b^-\,\right)}.
\label{Efield-0}
\end{equation}
where $J(t)$ is the electric  current density,
flowing through  the system. In view of the electroneutrality
condition, $J(t)$ is divergence-free, i.e., for the one-dimensional case
it can depend only on time. 

At this point, there are two ways to proceed. If a current 
generator is used and the current is fixed to be constant, $J(t)\equiv J$,
then Eq.(\ref{Efield-0}) determines $E(x,t)$ and thus 
Eqs.(\ref{electroneut1-0}-\ref{Efield-0}) together with the boundary
conditions Eqs.(\ref{bc1}-\ref{bc4})
form the closed set of equation to be solved. This is the 
case we shall treat in detail below.
 
Experimentally, it may be more convenient to maintain a {\em constant}  
voltage difference $U=V(L)-V(0)$ instead of a constant current. 
In that case, integrating Eq.(\ref{Efield-0}) yields
\begin{equation}
\int\limits_{-L/2}^{L/2}dx{ E(x,t)}=
\int\limits_{-L/2}^{L/2}dx\frac{J(t)}
{\lambda q\left(\,a^++a^-+b^++b^-\,\right)}=- U
\label{Efieldint-0}
\end{equation}
and so $J(t)$ is given through $U$. Then, substituting $J(t)$
into Eq.(\ref{Efield-0}), one finds 
the scaled field $\lambda E$ 
\begin{equation}
\lambda E(x,t) = -\lambda U \,\frac{
\left(\,a^++a^-+b^++b^-\,\right)^{-1}}{\int\limits_{-L/2}^{L/2}dx\,
\left(\,a^++a^-+b^++b^-\,\right)^{-1}}
\label{Efield-1}
\end{equation}
thus arriving at a closed set 
of Eqs.(\ref{electroneut1-0}-\ref{electroneut4-0}).

It should be noted that the constant $U$ and constant 
$J$ cases are equivalent provided only the 
stationary states are considered. Indeed, the concentrations are time 
independent in the stationary state and, consequently, it follows
from Eq.(\ref{Efieldint-0}) that constant $U$ implies that 
$J$ is independent of time, thus we returned to the constant 
current case. One cannot give, however, a simple relation
between $U$ and $J$ since, as can be seen from Eq.(\ref{Efieldint-0}), 
the proportionality constant depends on the stationary state reached
for a given $U$ or $J$.

\section{Stationary state}
\label{stationary}

The equations for the stationary solution $n_i(x,t)\equiv n_i(x)$ 
are obtained by setting the time derivatives to zero in Eqs.(\ref{electroneut1-0}-\ref{electroneut4-0}).
To make the equations more transparent one can
introduce dimensionless quantities by
measuring $x$ in units of $L$ and the concentrations $n_i$ 
in units of $a_0$. Then,
substituting Eq.(\ref{Efield-0}) with $J(t)=J$ into Eqs.(\ref{electroneut1-0}-\ref{electroneut4-0}) (i.e. considering the
constant current case), we obtain
\begin{eqnarray}
(a^-)^{''}
+J_0\left(\frac{a^-}{a^++a^-+b^++b^-}\right)^{'}-k_0a^-b^+=0\quad
\label{statio-1}\\
(b^+)^{''} -J_0 \left(\frac{b^+}{a^++a^-+b^++b^-}\right)^{'}-k_0a^-b^+=0\quad
\label{statio-2}\\
(a^+)^{''}-J_0\left(\frac{a^+}{a^++a^-+b^++b^-}\right)^{'}=0\quad
\label{statio-3}\\
(b^-)^{''}+J_0\left(\frac{b^-}{a^++a^-+b^++b^-}\right)^{'}=0\quad
\label{statio-4}
\end{eqnarray}
where prime denotes the spatial derivatives, and
\begin{equation}
J_0=\frac{JL}{qDa_0} \quad , \quad k_0=\frac{ka_0L^2}{D}\, .
\end{equation}

In order to solve the above equations the 
rescaled current $J_0$ and  rate constant 
$k_0$, and the boundary condition $b_0/a_0$ should be given. 
In principle, we could 
proceed then by numerically solving the equations. 

We would like,
however, to find first some analytical estimates of $w(J_0)$. 
For this purpose, we begin by
considering a symmetric setup where $a_0=b_0$ and
the $k_0\to \infty$ limit is taken. In this case, 
$J_0$ remains the only control parameter, and the 
reaction zone becomes pointlike. 
The limit is nevertheless relevant. First, because
the real reaction zones are narrow, in general. Second, because
the analytical estimates of the sensitivity of concentration profiles 
to small currents can be used to estimate the 
width for $k_0$ finite but large. Once the phenomenological 
estimate of $w(J_0)$ is obtained, we carry out numerical 
investigations as well to judge the accuracy of the phenomenology. 
In a final step, the results will be generalized 
to the $a_0\not= b_0$ case.

\section{Symmetric setup ($a_0=b_0$)}

\subsection{Point reaction zone (infinite reaction rate)}

The reaction zone 
is pointlike for $k_0\to\infty$, and it follows from the symmetry 
of the $a_0=b_0$ setup that 
the position of the reaction zone is at $x_f=0$. The concentrations 
of the reacting 
ions satisfy $a^-(x>0)=0$ and $b^+(x<0)=0$ and so we have
two sets of equations for the two sides of the front.
We begin by considering the left-hand side ($x<0$) and obtain 
the concentration on the right-hand side by symmetry, namely
$b_{x>0}^+(x)=a_{x<0}^-(-x)$, $b_{x>0}^-(x)=a_{x<0}^+(-x)$ and $a_{x>0}^+(x)=b_{x<0}^-(-x)$.

The equation for $x<0$ are found by setting $b^+=0$ in 
Eqs.(\ref{statio-1},\ref{statio-3},\ref{statio-4}):
\begin{eqnarray}
(a^-)^{''}
+J_0\left(\frac{a^-}{a^++a^-+b^-}\right)'&=&0\quad
\label{sinf-1}\\
(a^+)^{''}-J_0\left(\frac{a^+}{a^++a^-+b^-}\right)'&=&0\quad
\label{sinf-3}\\
(b^-)^{''}+J_0\left(\frac{b^-}{a^++a^-+b^-}\right)'&=&0 \quad .
\label{sinf-4}
\end{eqnarray}

The boundary conditions to the above equations 
at $-L/2$ [see Eqs.(\ref{bc1},\ref{bc2})] 
do not change while, at the reaction zone, the following
boundary conditions must be used
\begin{eqnarray}
a^-(0)&=&0\,,\label{a-=0}\\
b^-(0)&=&a^+(0)\,,\label{a+=b-}\\
\left.(a^+)'\right|_{x=0}&=&-\left.(b^-)'\right|_{x=0}
\quad .\label{deriv=}
\end{eqnarray}
The first equality follows from the infinite rate constant. The 
second one is the electroneutrality condition ($a^+-a^--b^-=0$) 
employed at the reaction zone. Finally, the third condition 
follows from the requirement that, for the counterions, not only
their concentrations but also their derivatives should be continuous
across the reaction zone. 

The first step in solving the equations for $x<0$ 
is the application of
the electroneutrality condition ($a^+-a^--b^-=0$)
to Eq.(\ref{sinf-3}). It yields $(a^+)^{''}=0$, thus resulting
in the following solution for $a^+$:
\begin{equation}
a^+(x)=1+A\left(2x+1\right)
\label{a+kinf}
\end{equation}
where we used the boundary condition at $x=-L/2$. The 
integration constant $A$ remains undetermined at this stage.

Having the solution for $a^+(x)$, and using the electroneutrality 
condition in Eq.(\ref{sinf-1}), we find now that $a^-$ satisfies
the following equation
\begin{equation}
 (a^-)'+\frac{J_0}{2}\frac{a^-}{1+A(2x+1)}=I_1
\label{s-1}\\
\end{equation}
where  $I_1$ is another integration constant.

The general solution of equation~(\ref{s-1}) 
reads
\begin{equation}
 a^-(x)=C_1[1+A(2x+1)]+C_2[1+A(2x+1)]^{\Delta} \, ,
\label{aminus}
\end{equation}
and the boundary conditions provide  
$C_1$ and $C_2$
\begin{eqnarray}
 C_1&=&-\frac{(1+A)^{\Delta-1}}{1-(1+A)^{\Delta-1}}\\
 C_2&=&\frac{1}{1-(1+A)^{\Delta-1}}
\end{eqnarray}
with $\Delta=-J_0/4A$.

Once $a^-$ and $a^+$ are known, $b^-$ is obtained from the
electroneutrality condition $b^-=a^+-a^-$. Thus what remained 
is to determine the integration constant $A$. We can find $A$
by combining the electroneutrality condition with the 
boundary condition Eq.(\ref{deriv=}) to arrive at 
\begin{equation}
\left.(a^-)'\right|_{x=0}=2\left.(a^+)'\right|_{x=0} \, .
\label{deriacross}
\end{equation}
Substituting now Eqs. (\ref{a+kinf}) and (\ref{aminus}) into 
Eq.(\ref{deriacross}), we obtain $A$ as the solution of the 
following relation:
\begin{equation}
 \left(1-\frac{J_0}{4A}\right)(1+A)^{-(1+\frac{J_0}{4A})}=2\;.
\label{AofJ}
\end{equation}

We have thus determined all the concentration profiles for $x<0$ and,
as mentioned above, the profiles for $x>0$ can be obtained 
from symmetry considerations. 

\subsection{Reaction rate and the slope of $a^-$ at $x=0$}
A simple question one might ask about the $k_0\to\infty$ case is 
the following: How does the rate of reaction changes when the current is 
switched on? In order to calculate this quantity, the slope
$(a^-)'$ needs to be evaluated at the reaction zone ($x=0$). As
we shall see, the same 
slope will also be important in the next
Subsection \ref{Pheno} where a phenomenological
estimate of the width $w(J)$ will be carried out.

The derivative $\left.(a^-)'\right|_{x=0}$ can be calculated 
by noting from Eq.(\ref{deriacross}) that $\left.(a^-)'\right|_{x=0}=2\left.(a^+)'\right|_{x=0}$
and that  $(a^+)'=2A$ follows from Eq.(\ref{a+kinf}). 
As a result, we arrive at
\begin{equation}
\left.(a^-)'\right|_{x=0}=4A \, .
\label{dam}
\end{equation}
The rate of reaction is the flux of the reacting ions $j$ 
into the reaction zone. It is 
the number of $A^-$ ions reaching $x=0$ in unit time, and is obtained 
as
\begin{equation}
j(J_0)=-\left.(a^-)'\right|_{x=0} - 
\left.\frac{J_0a^-}{a^++a^-+b^-}\right|_{x=0} \quad .
\label{currplusz1}
\end{equation}
where we note that 
$j$ is also scaled quantity, i.e. it is measured in units of $Da_0/L$. 
Since $\left.a^-\right|_{x=0}=0$, the second term on the 
right-hand side is zero. The first one is obtained from Eq.(\ref{dam}),
thus yielding
\begin{equation}
j(J_0)=-\left.(a^-)'\right|_{x=0}= -4A(J_0)
\label{currplusz2}
\end{equation}
where $A(J_0)$ is given by solving Eq.(\ref{AofJ}).

Analytic expressions of $j(J_0)$ can be developed for small $J_0$
since we can expand 
the solution of Eq.(\ref{AofJ}) in powers of $J_0\ll1$.
To first order in $J_0$, we find
$A(J_0)=-1/2+(1-\ln2)J_0/4$ and consequently
\begin{equation}
j(J_0)=-\left.(a^-)'\right|_{x=0}=2-(1-\ln2)J_0 \quad .
\label{dam1}
\end{equation}
Introducing the scaled flux $\tilde j_0$ of $A^-$ ions 
in the absence of electric 
current ($\tilde j_0=2$), we can write the expansion as 
\begin{equation}
j(J_0)=\tilde{j_0}\left(1-\alpha \frac{J_0}{\tilde{j_0}}\right) \quad .
\label{dam4}
\end{equation}
where $\alpha=1-\ln{2}>0$.

Let us consider now the sign of the first order contribution. 
Note that $J_0\sim J$ and it follows from
Eqs.(\ref{electroneut1-0}-\ref{electroneut4-0}) that 
$J<0$ means that the negative (positive) ions are driven 
to the right (left). In this case, the reacting ions
are driven towards the reaction zone and 
the first order correction is positive and, consequently, 
the reaction rate increases [$j(J_0)>\tilde{j_0}$]. We shall call 
the $J<0$ electric current as {\it forward 
current}. The reagents are driven away from the reaction zone 
in the opposite case ($J>0$, {\it backward current}) and, as expected,
the reaction rate decreases [$j(J_0)<\tilde{j_0}$]. 

An important point to 
recognize here is that the change in the reaction rate does not come 
directly from the drift term in the flux of particles, it comes 
indirectly from the change in the diffusive flux i.e. from 
the change in the slope of the reaction profiles near 
the reaction zone. We shall see a similar effect when calculating 
the width of the reaction zone.

In closing this section, we display the diffusive flux 
$-D(a^-)'$ in terms of the original variables 
\begin{equation}
\left.-D(a^-)'\right|_{x=0}=2\frac{a_0D}{L}+(1-\ln2)\frac{J}{q} =
j_0-\alpha \frac{J}{q}\quad .
\label{dam2}
\end{equation}
where $j_0=2a_0D/L$ is the unscaled diffusive flux in the absence 
of electric current. 
We shall need the above expression in the phenomenological 
arguments of the next section which
would be less transparent using the scaled variables.

\subsection{Finite reaction rate: Phenomenological considerations and numerical solutions}
\label{Pheno}

For the limiting case of an infinite rate constant ($k_0\to \infty$), 
we have analytical solutions for the concentration profiles for
arbitrary $J_0$. The physically relevant case, however, is the one 
with finite $k_0$ where there is little hope to find exact
solutions. For this case, we developed phenomenological 
considerations which have their roots in the success of similar 
arguments for simpler cases \cite{zol,Redner1}, and we expect it 
to be performing well at least for small currents. 
As we shall see, the validity of the phenomenological argument is 
supported by numerical integration of the original equations.

The main idea is to use the balance equation for the reacting ions,
namely to equate the number of reactions per unit time to
the flux of reacting ions towards the reaction zone. To do this, 
one needs the values of the concentrations and their derivatives. 
They are estimated by assuming that the concentration profiles 
are smooth functions and, for finite but large $k_0$, their
slopes in the reaction zone can be approximated by the 
$k_0\to \infty$ values.  
 
The balance equation can be obtained by
integrating eq.(\ref{statio-1}) through the reaction zone 
i.e. from $-w$ to $w$
\begin{equation}
\left.-D\frac{da^-}{dx}\right|_{-w}-\frac{J}{q}
\left.\frac{a^-}{a^++a^-+b^++b^-}\right|_{-w}=k\overline{a^-}\cdot
\overline{ b^+}w\,.
\label{w-estimate1}
\end{equation}
Here the upper bars represent the spatial averages over the 
reaction zone and, in the spirit of mean-field approximation, 
the average of the 
product $\overline{ a^-b^+}$ has been replaced by the product 
of averages $\overline{a^-}\cdot\overline{b^+}$. Furthermore, the 
contribution from the upper 
limit of the integrals on the left-hand side have been neglected
since $a^-$ and $(a^-)'$ become zero at the right end of the reaction
zone [$a^-(w)\approx 0$, $(a^-)'(w)\approx 0$].

The first term on the left hand side is the diffusive flux
which has to be evaluated to first order in 
$J$. Our approximation consists in replacing the slope $(a^-)'$ 
at $x=-w$ by $(a^-)'$ at $x=0$  
and use the expression obtained in the $k\to\infty$ limit 
given by eq.(\ref{dam2}).
The second term can be approximated by replacing the denominator by
$a_0$ (it is exact in the $k\to \infty$ and $J\to 0$ limit). As we shall
see, this term turns out to be negligible thus the details are 
immaterial. As a result, we find
\begin{equation}
j_0-\alpha \frac{J}{q}-\frac{J}{q a_0}\left.a^-\right|_{-w}=k\overline{a^-}\cdot
\overline{ b^+}w\,
\label{w-estimate2}
\end{equation}
The values of $\left.a^-\right|_{-w}$ and $\overline{a^-}\approx
\overline{ b^+}$ can be estimated by noting that $a^-(w)\approx 0$ 
thus the function at $x\approx 0$ is approximately 
given as $(a^-)'(-w)$. Thus, we have
\begin{equation}
\left.a^-\right|_{-w}\approx \overline{a^-}\approx
\overline{ b^+} \approx -\left.(a^-)'\right|_{x=0}w\approx \frac{w}{D}(j_0-\alpha \frac{J}{q})\,
\label{w-estimate3}
\end{equation}
where we neglected multiplicative factors of order 2, and 
used again Eq.(\ref{dam2}) for evaluating $\left.(a^-)'\right|_0$.

Substituting the above expressions into (\ref{w-estimate2}) we obtain
\begin{equation}
j_0-\alpha \frac{J}{q}-\frac{Jw}{qa_0D}j_0
\approx k\frac{w^3}{D^2}\left(j_0-\alpha \frac{J}{q}\right)^2\;.
\label{w-estimate4}
\end{equation}
For zero electric current ($J=0$), the solution of this equation is 
\begin{equation}
w_0\approx\left(\frac{D^2}{kj_0}\right)^{1/3},
\label{w0}
\end{equation}
which is a result obtained in a study of the reaction zone 
in case of neutral reagents~\cite{redner}.

Expanding now the solution of Eq.(\ref{w-estimate4}) to first order 
in $J/qj_0$, one finds
\begin{equation}
w=w_0\left[1+\left(\frac{\alpha}{3} - \frac{2w_0}{3L}\right) \frac{J}{qj_0}\right]
\quad .
\label{w3}
\end{equation}
Since, for the usual situation of large $k$, we have $w_0\ll L$, the 
last term on the right-hand side (whose origin can be traced back to 
the drift term in the flux of the ions) can be 
neglected. Thus we arrive at the 
final form for the width of the reaction zone:
\begin{equation}
\frac{w}{w_0}=1+\frac{1-\ln2}{3} \, \frac{J}{qj_0}
\quad .
\label{w4}
\end{equation}
This is the central result of the paper. It tells us that the width
of the reaction zone decreases for forward currents ($J<0$) while 
it increases for current of opposite polarity. 

\begin{figure}[htbp]
\includegraphics[width=7cm,angle=0,clip]{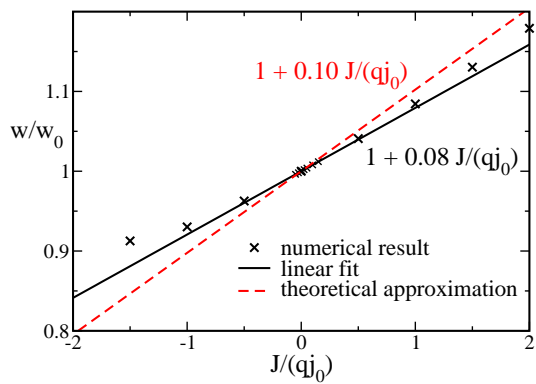}
\caption{(Color online) The small-current expansion of the width 
from numerical integration of the Eqs.(\ref{sinf-1}-\ref{sinf-4}) is 
compared to the phenomenological theory of Subsection \ref{Pheno}.  
The width is scaled by its value at $J=0$ while the current is 
scaled by $qj_0$ where $j_0$ is the particle flux at $J=0$.} 
\label{width}
\end{figure}

The numerical factor 
in front of  $J/qj_0$ is of course a question and this is why
we carried out the numerical integration of the differential 
equations. The equations are simple and the numerics does not pose
any problem. Fig.\ref{width} shows the comparison of the numerical 
results with the phenomenological theory. 
The agreement is surprisingly good in view of the 
simplicity of the phenomenological arguments we used.
Thus our assumptions entering the derivation of $w(J)$ in the small $J$
and large $k$ limits appear to be justified.

\section{Asymmetric setup}
\label{asym}

We would like to control the width and so, it is useful to consider
a more general case provided by asymmetric
boundary condition $a_0\not=b_0$ since it
introduces a new control parameter 
\begin{equation}
Q=\frac{b_0}{a_0} \quad .
\end{equation}
The study of this case follows the steps of the symmetric case. 
First, the point like reaction zone ($k_0\to \infty$) is solved
exactly, then the phenomenological consideration are repeated, and
finally, the numerical simulations are used to check the validity
of the phenomenology.

As before, for $k_0\to \infty$, the left and the right sides of 
the front can be treated separately.
For the left hand side ($x<x_f$) one can use the 
equations~(\ref{sinf-1}), (\ref{sinf-3}) and (\ref{sinf-4}) while
for the right hand site ($x>x_f$) the equations are as follows
\begin{eqnarray}
(b^+)^{''}
-J_0\left(\frac{b^+}{b^++b^-+a^+}\right)'=0\quad
\label{sinfr-1}\\
(a^+)^{''}-J_0\left(\frac{a^+}{b^++b^-+a^+}\right)'=0\quad
\label{sinfr-3}\\
(b^-)^{''}+J_0\left(\frac{b^-}{b^++b^-+a^+}\right)'=0\quad ,
\label{sinfr-4}
\end{eqnarray}
and the electroneutrality condition for $x>x_f$ reads
\begin{equation}
a^++b^+-b^-=0\quad .
\end{equation}

These equations are solved with the following boundary conditions 
for the dimensionless quantities. At the outer boundaries we have 
$a^\pm(-1/2)=1$, $b^-(-1/2)=0$, $a^+(1/2)=0$, and $b^\pm(1/2)=Q$. 
At the front ($x=x_f$), one has $a^-(x_f)=0, ~b^+(x_f)=0$ and $a^+(x_f)=b^-(x_f)$. 
Finally, the conditions for the slopes at $x=x_f$ are:
\begin{eqnarray}
 \left.(a^-)'\right|_{x_f^-}&=&-\left.(b^+)'\right|_{x_f^+}\\
\left.(a^+)'\right|_{x_f^-}&=&\left.(a^+)'\right|_{x_f^+}\\
\left.(b^-)'\right|_{x_f^-}&=&\left.(b^-)'\right|_{x_f^+}
\label{a-derivativenew}
\end{eqnarray}
where $x_f^-$ ($x_f^+$) denotes the left (right) side of the 
reaction zone. 

Since the solution of the above problem
is similar to the symmetric case (though the algebra
is much more tedious), we present only the main results. 

\begin{figure}[htbp]
\includegraphics[width=7cm,angle=0,clip]{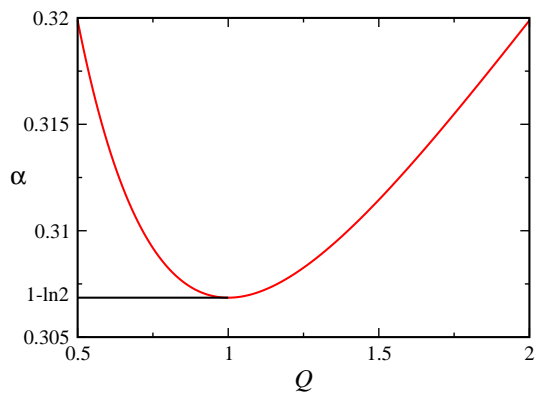}
\caption{(Color online) $Q$ dependence of the parameter $\alpha$.} 
\label{fig4}
\end{figure}

Due to asymmetry, the position of the front $x_f$ shifts from $x=0$
and one finds that it is given by
\begin{eqnarray}
 x_f&=&\frac{1-Q}{2(1+Q)}+\frac{J_0}{2Q(1+Q)^2} \times \\
&&\left[Q-Q^2+\ln\left(\frac{1}{1+Q}\right)-
Q^3\ln\left(\frac{Q}{1+Q}\right)\right] \nonumber\;.
\label{newfront}
\end{eqnarray}
An important quantity for the phenomenological arguments is 
the slope of the concentration of the reacting ion $a^-$ on the
left-hand side of $x_f$. It is found to be 
\begin{equation}
 \left.(a^-)\right|_{x_f}=-(1+Q)+\alpha(Q) J_0
\label{newslope}
\end{equation}
with
\begin{equation}
 \alpha(Q)=1+\frac{1}{2Q}\ln\left(\frac{1}{1+Q}\right)+
\frac{Q}{2}\ln\left(\frac{Q}{1+Q}\right)\, .
\label{alpha}
\end{equation}
Fig.~\ref{fig4} shows 
the dependence of $\alpha$ on $Q$ and one can observe that 
$\alpha(Q)$ has a minimum at $Q=1$ i.e. in the symmetric setup. As expected the expression ($\ref{alpha}$) is invariant under the exchange of the boundary 
concentration values, $a_0$ and $b_0$, i.e.~ $\alpha(Q)=\alpha(1/Q)$.

For finite reaction rate, one can follow the steps of the 
phenomenological arguments of the symmetric case, and
calculate the first order expansion of $w(J)$ in $J/(qj_0)$ where
$j_0$ is given now by 
\begin{equation}
 j_0=(a_0+b_0)D/L \, .
\end{equation}
The final result is  
\begin{equation}
 \frac{w}{w_0}=1+\frac{\alpha(Q)}{3}\frac{J}{qj_0}
\label{neww}
\end{equation}
suggesting that the width of the reaction zone is more effectively 
reduced when the system is asymmetric.  

The predictions of the phenomenological 
arguments can again be compared with the results of a numerical 
integration of the differential equations. The agreement is 
qualitatively similar to the symmetric case and the tendency of 
an increasing impact of the imposed current 
on the reaction front width when enhancing the asymmetry 
has been verified.

\section{Considerations about the width of a moving front}

Although our motivation for this work 
comes from controlling precipitation patterns which emerge 
in the wake of a  {\it moving} 
reaction-diffusion front \cite{ourprl, encoding}, 
as a first step, we considered the controllability of the 
width in a simpler {\it stationary} state. Clearly, it should be now
clarified to what extent the results are generalizable to 
the case of moving fronts. 

The first step in the generalization is the
realization that diffusion 
fronts slow down with time (their velocity is proportional 
to $1/\sqrt{t}$). This means that the system enters a 
quasistationary regime where stationary state considerations
usually yield correct result. As a relevant example here, 
let us examine the time 
evolution of the width of the reaction zone $\tilde w_0(t)$ emerging 
when the reagents are initially separated and no current 
is present [$a(x<0,t=0)=a_0$, $b(x<0,t=0)=0$, $a(x>0,t=0)=0$,
$b(x>0,t=0)=b_0<a_0$, $L\to\infty$]. In this case, the front 
moves diffusively and the width grows with time
as \cite{zol} $\tilde w_0(t)\sim t^{1/6}$. This time-evolution 
can be derived easily from the stationary state 
result\cite{redner} $w\sim 
(D^2/kj_0)^{1/3}$ [see eq.(\ref{w0})] by the following argument.
The width
of depletion zone (the region where the concentration profiles 
are significantly different from their initial values) grows
with time as $W_d(t)\sim\sqrt{Dt}$. Then the diffusive 
current towards the reaction zone is obtained as
$j_0(t)\sim D(a_0+b_0)/W_d(t)\sim t^{-1/2}$, and
one finds the desired time-evolution as
\begin{equation}
\tilde w_0(t)\sim \left(\frac{D^2}{kj_0(t)}\right)^{1/3}\sim t^{1/6}\, .
\end{equation}

The second property of the front needed for the generalization
is that the motion of the front remains diffusive 
even if the ionic nature of the reagents are taken into account 
and a small current is switched on \cite{bena}. Assuming 
that the diffusive nature of the front implies 
quasistationarity, one can go through the estimates of the 
various terms in the reaction-diffusion equations provided 
those terms are restricted to the neighborhood (the region 
of depletion zone) of the reaction front. Using the 
quasistationary time evolution of the diffusive current 
$j_0\sim t^{-1/2}$, one finds then that the neglected time 
derivatives are indeed small provided $\tilde w(t)\ll W_d(t)$. 
Furthermore, following along the same steps which lead to
the estimates of $w/w_0$ in the small electric current 
limit [see the derivations of eqs.(\ref{w4})
and (\ref{neww})
in Sec.\ref{Pheno} and Sec.\ref{asym}], one finds
\begin{equation}
\tilde w(t)\approx \tilde w_0(t)
\left[1+\tilde\alpha\frac{J}{qj_0(t)}\right]
\label{wmove}
\end{equation}
where $\tilde \alpha$ is a numerical constant of the order one.
An important conclusion to be drawn from the above expression 
is that the direction 
of the current has the same effect as before. Namely, the forward 
current decreases the width while the backward current increases it.
 
Unfortunately, the decrease of $j_0\sim t^{-1/2}$ makes it clear
that the expansion of $\tilde w$ in $J/qj_0$ cannot be valid 
for large times. It is valid only at intermediate
times when the quasistationary regime has already set in but 
the electric current is still smaller than the diffusive current.

Clearly, we cannot claim here that the width of a moving front 
in the presence of a current has been understood. We believe, 
however, that the solution should have the properties
described above in a well-defined intermediate-time regime. 
Finally, we should also note that, in the control
of precipitation patterns \cite{ourprl}, one employs actually
time-dependent currents and one has some freedom 
in the design of those current. Consequently,
the effects of variable
currents on the width, and the use of the 
freedom in the choice of current 
to control the width provide additional challenges
which should be treated once we arrived at basic understanding 
of the simpler problems.

\section{Discussion and Final remarks}

We have studied 
the effects of an external electric current 
(or potential difference) on the width of the reaction zone in an 
irreversible $A^- +B^+ \to C$ reaction-diffusion process.
The aim was to find a flexible control of the width since this 
would help in the attempts of downscaling the patterns generated 
by reaction-diffusion processes. 

Unfortunately, we have to conclude that the width of the reaction zone is 
not very sensitive to the presence of an externally imposed electric current or
potential difference. In the symmetric setup for a typical values $J/q j_0$ of order $-1$, the reduction of the 
width compared to the zero-current case is of the order of 10\% at best 
(see Fig.\ref{width}) and, furthermore, even a strong 
asymmetry in the boundary conditions ($Q\to \infty$) allows only
a reduction of the order of 20\% [see Eqs.(\ref{alpha},\ref{neww})].

Clearly, the new attempts at control should start now with the 
reexamination of $w_0$. In order to see the problem, 
let us write Eq.(\ref{w0}) 
in terms of the characteristic concentration and lengthscale
of the system
\begin{equation}
w_0\approx\left(\frac{D^2}{kj_0}\right)^{1/3}\approx
\left(\frac{D}{k}\cdot\frac{L}{a_0}\right)^{1/3}.
\label{w0-2}
\end{equation}

The rate constant $k$ and the diffusion constant $D$
cannot be controlled effectively. So, what remains to consider 
is the gradient of concentrations, $a_0/L$, and one can see 
from Eq.(\ref{w0-2}), that small width is obtained by 
large concentration gradients. 
Actually, the wet stamping method \cite{Grzy1,Grzy2} 
employs large gradients ($a_0\approx 1M$ and 
$L\approx 100\mu$) to achieve patterns with features 
at submicron scales. The problem with wet stamping, however, is 
that the patterns are generated by skillfully preparing the 
boundary conditions and thus the method runs into difficulties with 
the task of building three-dimensional patterns. Nevertheless, 
the wet stamping points towards a possible solution 
of our problem. Namely, we
would have to design reaction-diffusion processes which can 
support large concentration gradients in a 
reaction zone whose motion can be controlled. 
Of course, this appears to be a nontrivial task.

Although the diffusion constant $D$ cannot be controlled,
one can still think about considering materials where
$D$ is significantly different from those of the ions in
gels ($D_{gel}\approx 10^{-8}m^2/s$). In solid state systems, 
like for example glasses, the diffusion coefficients differ 
strongly from the typical values in gelatine materials
(typically $D_{solids}\approx 10^{-18}m^2/s$) \cite{Halle}.
Accordingly, the zero-current width can be
extremely small, since $w(0)\sim D^{2/3}$ (see Eq. (61)).

The study of glassy systems is also interesting
since the dynamics of reactants may become subdiffusive
in a glassy matrix.
For subdiffusive systems we have $D\to 0$ and, as recent
studies have shown, the subdiffusive reaction-diffusion dynamics leads to striking changes
in the concentration profiles of the reacting ions
\cite{Sokolov}. Thus, combining subdiffusivity with
the use of electric currents appears to be a promising
way of producing controlled reaction zones of very
narrow (submicron) width.

\acknowledgments

This research has been partly supported by the 
Swiss National Science Foundation and
by the Hungarian Academy
of Sciences (Grants No.\ OTKA K 68109).


\end{document}